\documentclass[paper,12pt]{article}
\pdfoutput=1

\usepackage[pdftex,bookmarks,colorlinks]{hyperref}
\usepackage[pdftex]{graphicx}

\usepackage{mathrsfs,mathcomp}
\usepackage{fancyhdr}
\fancypagestyle{plain}{
\fancyhead[CO]{\footnotesize{Candidate video for the 28th Annual Gallery of Fluid Motion\\ 63rd Annual Meeting of the American Physical Society, Division of Fluid Dynamics \\Long Beach, CA, 21-23 Nov 2010}}
}
\begin{document}

\title{Avalanche of particles in evaporating coffee drops}

\author{\'Alvaro G. Mar\'in\footnote{a.g.marin@utwente.nl}, Hanneke Gelderblom,\\
Jacco Snoeijer, Detlef Lohse\\
\emph{Physics of Fluids, University of Twente}\\
%\footnote{h.gelderblom@utwente.nl}
%\small{Poster for Lorentz Center's Workshop Capillary shaping of solutes, May 17th-21st 2010}
}

%\begin{affiliations}
%\item Physics of Fluids, University of Twente
%\end{affiliations}

\maketitle

\begin{abstract}

The pioneering work of Deegan et al. [Nature 389, (1997)] showed how a drying sessile droplet suspension of particles presents a maximum evaporating flux at its contact line which drags liquid and particles creating the well known \emph{coffee stain ring}. In this \emph{Fluid Dynamics Video}, measurements using micro Particle Image Velocimetry and Particle Tracking clearly show an avalanche of particles being dragged in the last moments, for vanishing contact angles and droplet height. This explains the different characteristic packing of the particles in the layers of the ring: the outer one resembles a crystalline array, while the inner one looks more like a jammed granular fluid. Using the basic hydrodynamic model used by Deegan et al. [Phys. Rev. E 62, (2000)] it will be shown how the liquid radial velocity diverges as the droplet life comes to an end, yielding a good comparison with the experimental data.

\end{abstract}

\section{Technical Details of the submitted video}

A single experiment is shown in the included video (\href{http://stilton.tnw.utwente.nl/people/alvaro/Avalanche_gfm_512x384.avi}{low quality}, \href{http://stilton.tnw.utwente.nl/people/alvaro/Avalanche_gfm_1024x768.avi}{higher quality}): an evaporating droplet of about $3\mu l$ containing fluorescent polystyrene particles of $1 \mu m$ in a concentration of $0.2\% w/w$ ($\sim 7x10^9 part/cm^3$) evaporates at $23\tccelsius$ and controlled humidity of $30\%$ on a thin glass slide. Visualization is performed from side view and bottom view simultaneously. For the side view, a x10 long distance microscope and a high definition digital video camera is employed. The images are then processed with a self-made MATLAB algorithm to obtain contact angle, volume and radius of the droplet in time. Bottom view is performed with an inverted microscope, using x40 magnification and a intensified PCO sensicam camera and focusing at the first observable layer of particles close to the glass slide. A self-made MATLAB $\mu$PIV algorithm is then used to process the images and extract the velocity field in time.

For more information about this work, we would like to invite you to assist to the presentation from Hanneke Gelderblom at the 63rd Annual American Physics Society Meeting (Division of Fluid Dynamics) in Long Beach, CA (Abstract CS.00004, Nov 21st, 2010).

\bibliographystyle{plainnat}

\end{document}